\documentclass{elsart}
\input{psfig}

\begin{document}

\begin{frontmatter}
\title{Measurements and Simulation Studies of Piezoceramics for Acoustic Particle Detection}\footnote{\uppercase{T}his work is supported by the \uppercase{G}erman \uppercase{BMBF G}rant \uppercase{N}o. 05 \uppercase{CN}2\uppercase{WE}1/2.}

\author{K. SALOMON, G. ANTON, K. GRAF, J. H\"OSSL,}
\author{A. KAPPES, T. KARG, U. KATZ,}
\author{R. LAHMANN and C. NAUMANN}

\address{University of Erlangen \\
Erwin-Rommel-Str. 1, \\ 
91058 Erlangen, Germany\\ 
E-mail: karsten\_salomon@web.de}

\begin{abstract}Calibration sources are an indispensable tool for all detectors.  In acoustic particle detection the goal of a calibration source is to mimic neutrino signatures as expected from hadronic cascades. A simple and promising method for the emulation of neutrino signals are piezo ceramics. We will present results of measruements and simulations on these piezo ceramics.
\end{abstract}

\end{frontmatter}

\section{The Piezoelectric Effect and Signal Propagation in Water}
The active element of a transducer is a piezoelectric material. When an electric field is applied to the material, the polarized molecules are distorted in the electric field. This distortion causes the material to change its dimensions. This phenomenon is known as electrostriction or inverse piezoelectric effect. In addition, a permanently-polarized material such as lead zirconate titanate (PZT) 
produces an electric field when the material changes dimensions as a result of an imposed mechanical force. This phenomenon is known as the piezoelectric effect.
A piezoelectric material can be modeled by connecting Hooks law for anisotropic materials to the Gauss law for electrical displacement.
This is described by tensor equations:
\begin{equation}\label{piezoequ}
\partial_k(c_{iklm}S_{lm}+e_{lik}E_l)=\rho \ddot{u}_i\ ;\ \ \ 
\partial_i(e_{ilm}S_{lm}+\epsilon_{il}E_l)=0
\end{equation}
with $c_{iklm}$, $S_{lm}$, $e_{ikl}$, $E_l$, $\epsilon_{il}$, $\rho$ and $u_i$ representing the elasticity tensor, the strain tensor, the piezoelectric tensor, the electric field, the permitivity tensor, the density and the displacement\cite{HAllik}.

With the displacement response of a piezo due to an applied voltage (described in section~\ref{DispRes}) one can study how the motion of a piezo transforms into a pressure signal. For wavelengths larger than the dimension of the transmitter the resulting pressure is given by\cite{Landau}
\[p(\mathbf{r},t)=-\rho_0\frac{\partial_t^2 V(|\mathbf{r}|-t/c)}{4\pi r},\] 
where $\rho_0,r,t,V\approx Adx$ and $c$ are the density of the medium, the location, the time, the volume of the piezo, the speed of sound, $A$ is the area of the face of the piezo and $dx$ the displacement at the centre of the face of the piezo, respectively. Thus the pressure pulse sent by a transducer depends on the second derivative in time of the displacement.
\section{Mechanical Properties}
\begin{figure}[ht]
\psfig{figure={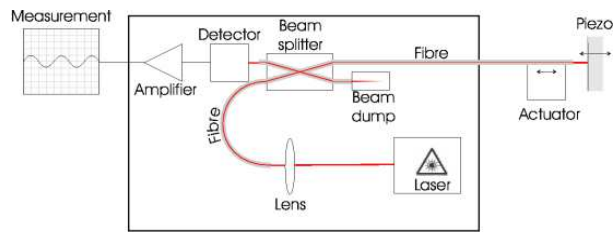},width=8cm}
 \caption{Schematic view of the interferometer setup}
  \label{figSchemaInterferometer}
\end{figure}

We have studied the mechanical behaviour of the sensors via direct measurement of the displacement of the surface of a piezo when a voltage is applied with a fibre coupled interferometer\cite{DRug,MDien} as shown in Fig.~\ref{figSchemaInterferometer}. 
Laser light ($\lambda$=780 nm) is coupled into an optical fibre. One light path reaches the end of the fibre after a 2x2 beam splitter. The fibre exit is prepared to behave like a mirror. This results in both reflection of the light back into the fibre (about 4\% for glass to air) and transmission. The face of the piezo is then positioned some micrometers away from the fibre exit. Multiple reflections between the fibre ending and the piezo\footnote{A glass plate vapour deposited with gold is glued to the piezo to increase reflections} lead to an interference. At the beam splitter light is guided to a photo diode, where the intensity of this light is measured after amplification. To calibrate the interferometer, an actuator is attached to the fibre ending. A resolution of $10^{-10}$m is achieved with this setup.
\label{DispRes}

\begin{figure}[ht]
\psfig{figure={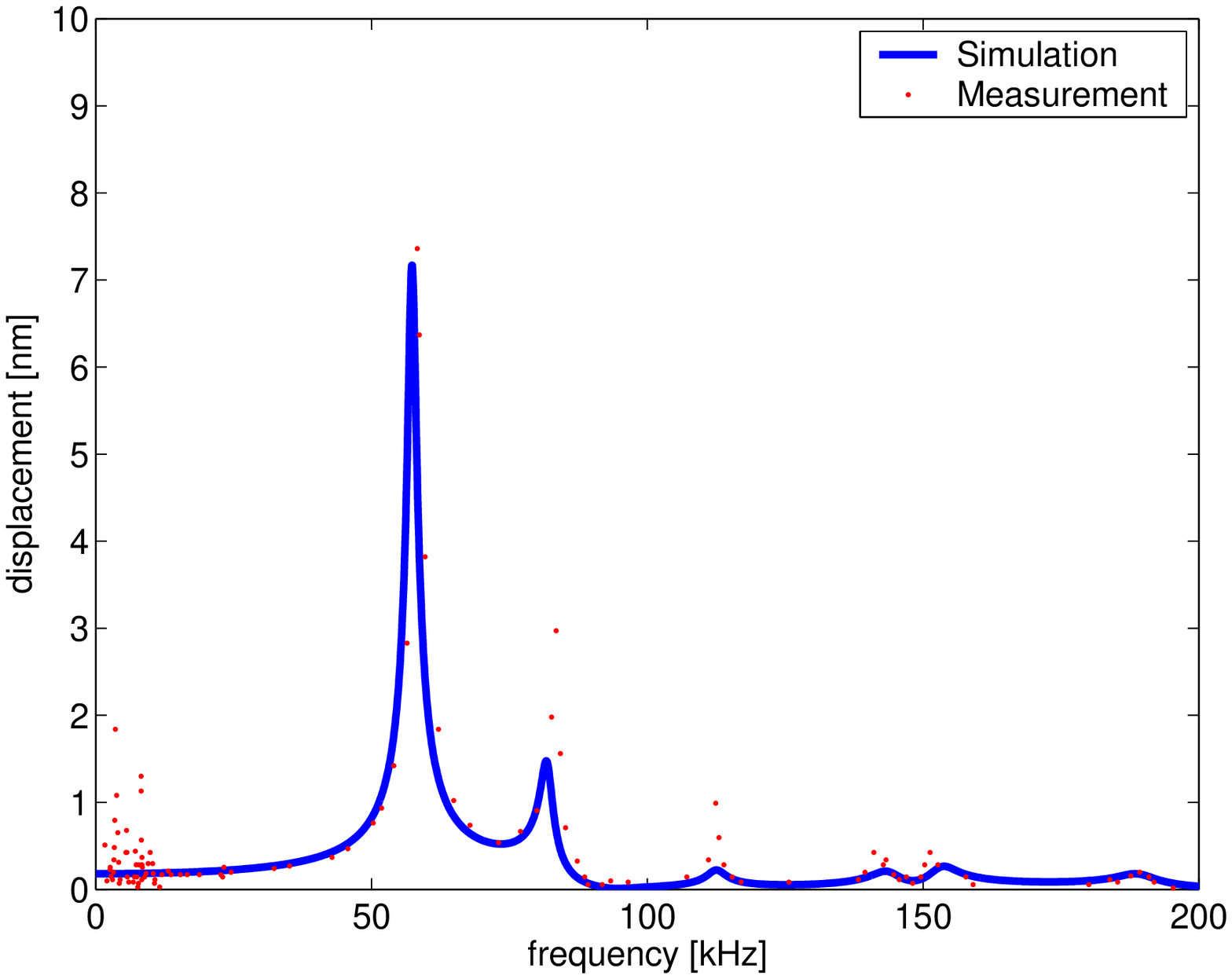},width=8cm}
 \caption{Displacement of a piezo disc r=12mm h=19.5mm}
  \label{figDisp}
\end{figure}

Using the finite element program CAPA\cite{RLerch} one is able to solve the piezoelectric Eq.~(\ref{piezoequ}) numerically. In addition CAPA uses a Rayleigh model to describe damping. The equation is solved in the frequency domain such that the differentials in time are transformed to algebraic equations.
Results of the comparison between simulation and measurement are shown in Fig.~\ref{figDisp}.
The positions of the resonances are in excellent agreement with the measurement. The amplitude of the displacement at the resonances differs from the measurement due to unknown damping parameters.

\begin{figure}[ht]
\psfig{figure={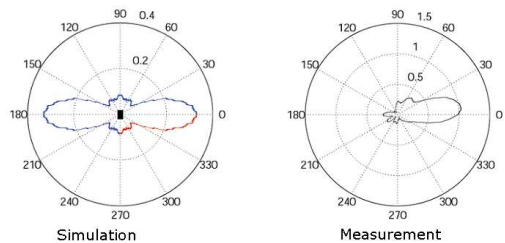},width=8cm}
 \caption{Direction Characteristics of a piezo disc in a.u.; Left: Simulation when sending (red: Simulation, blue: from simulation due to symmetry); Right: Measurement when receiving.  Asymmetry in measurement due to shadowing of the cast integral amplifier.}
  \label{figDirect}
\end{figure}

The propagation of sound in water is included in CAPA. From the resulting pressure field as a function of time in water one can extract the direction characteristics of the piezo by taking the local mean pressure field in the far field. (See Fig.~\ref{figDirect})		  
\section{Electrical Properties}
\begin{figure}[ht]
\psfig{figure={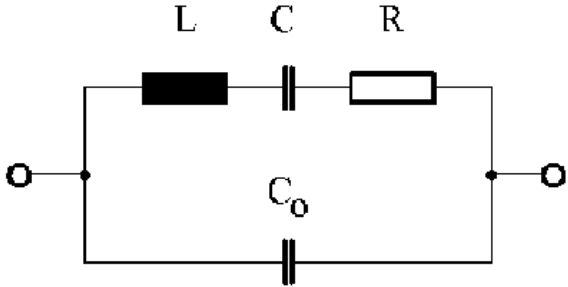},width=8cm}
 \caption{Equivalent circuit diagram for the first resonance of a piezo.}
  \label{figeCD}
\end{figure}

The first resonance and antiresonance of a piezo can be emulated with the equivalent circuit diagram shown in Fig.~\ref{figeCD} where the inductivity $L$, capacity $C$ and resistance $R$ correspond to the mass, stiffness and damping respectively. Further resonances can be emulated by additional parallel circuits. The total capacity in Fig.~\ref{figeCD} equals the capacity of a piezo. With this equivalent circuit diagram a fit to the measured impedance can be performed to extract $L,\ R,\ C$ and $C_0$.

\begin{figure}[ht]
\psfig{figure={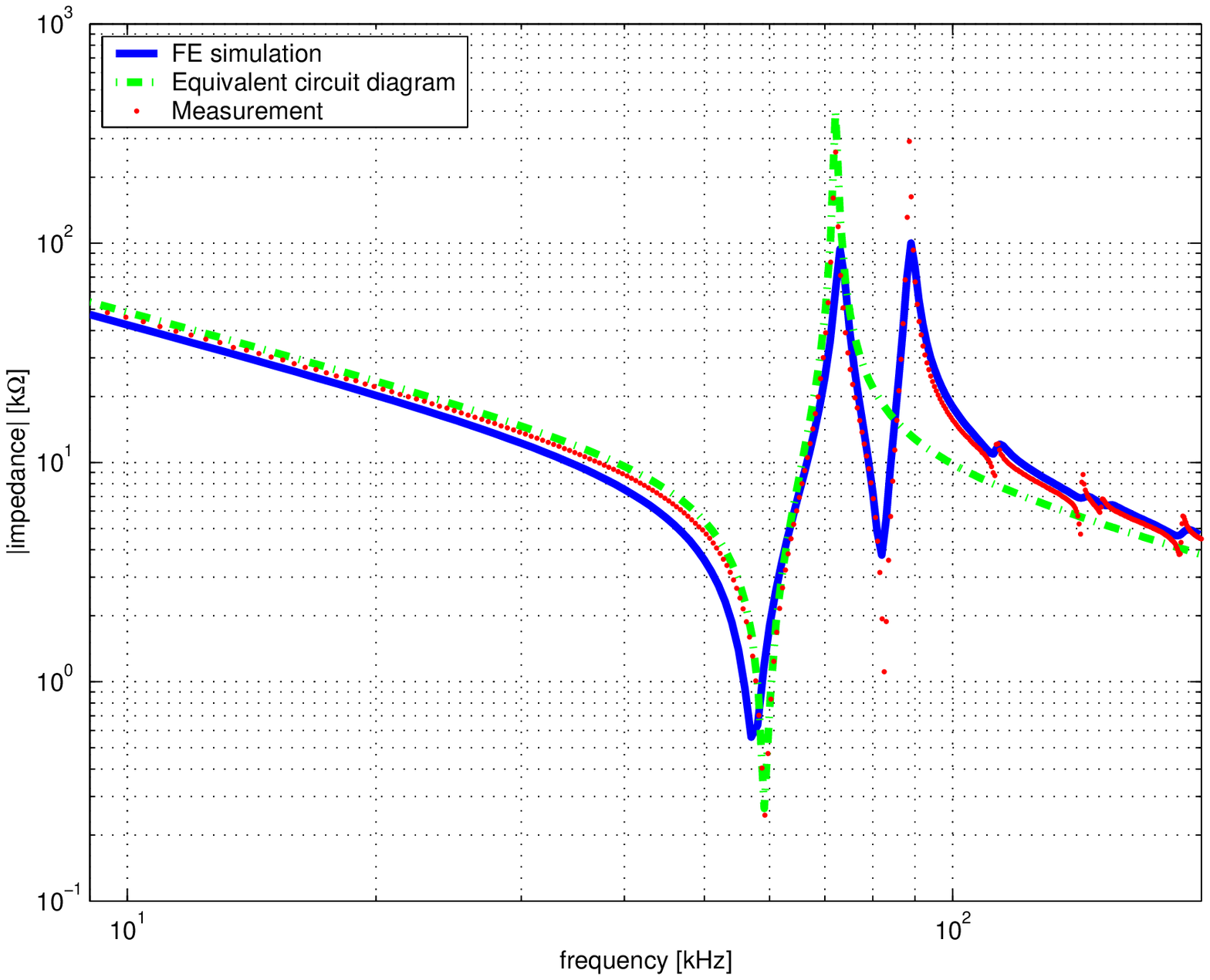},width=8cm}
 \caption{Impedance of a piezo disc r=12mm h=19.5mm}
  \label{figImp}
\end{figure}

	For the simulation of the impedance using CAPA a charge pulse  $Q(t)$
has been applied to the piezo and the voltage response to this charge was calculated. In the fourierspace the impedance is then given as:
 $Z(\omega)=\frac{\widetilde{U}(\omega)}{\widetilde{I}(\omega)}=\frac{\widetilde{U}(\omega)}{i\omega \widetilde{Q}(\omega)}$,
where $U,\ I\ $and $Q$ are the calculated voltage, the current and the applied charge, respectively. A comparison of the measurement with the simulation and the results from the fit of the equivalent circuit diagram to the measurement are shown in Fig.~\ref{figImp}.

The fit using the equivalent circuit diagram is in excellent agreement with the first resonance and antiresonance. The FE simulations fail to describe exactly the height of the resonances because of unknown damping parameters (even worse than for the displacement - see section~\ref{DispRes}).
\section{Summary and Conclusion}
We have shown that Finite Element Methods are a valuable tool to study piezoceramic sensors.
Comparing results of a piezo based on these methods with measurements obtained from an optic fibre interferometer yields very good agreement. Also the direction characteristics of the piezo are in good agreement with the simulation. The measured impedance can be very well reproduced using an equivalent circuit diagram on one side and using FE methods on the other. 
This detailed understanding of the piezo characteristics will enable us to both design the optimal sensors for acoustic particle detection and the transducers to calibrate these sensors.


\begin{thebibliography}{0}

\bibitem{HAllik} H. Allik and T. J. R. Hughes, {\it Int. J. Num. Meth. Eng.} {\bf 2}, 151-157 (1970).
\bibitem{Landau} L. D. Landau, {\it Hydrodynamik}, Akademie-Verlag Berlin.%

\bibitem{DRug} D. Rugar, H.J. Mamin and P. Guethner, {\it Appl. Phys. Lett.}  
{\bf 55}, 25 (1989).

\bibitem{MDien} M. Dienwiebel, et al. 
{\it Rev. Sci. Instrum.} {\bf 76}, 043704 (2005).

\bibitem{RLerch} R. Lerch {\it IEEE Transaction on UFFC} {\bf 37 (3)}, 233-247 (1990), ISSN 0885-3010.

\end{thebibliography}
\end{document}